\documentclass[twocolumn,showpacs,superscriptaddress,preprintnumbers,amsmath,amssymb,aps,prl]{revtex4}

\usepackage{CJK}

\usepackage{amsmath}
\usepackage{graphicx}
\usepackage{dcolumn}
\usepackage{color}
\usepackage{bm}
\usepackage{ulem}

\begin{document}
\begin{CJK*}{GB}{SongMT}
\CJKfamily{gbsn}

\title{Complete orbital angular momentum Bell-state measurement and superdense coding}

\author{Ling-Jun Kong}
\author{Rui Liu}
\author{Zhou-Xiang Wang}
\author{Yu Si}
\author{Wen-Rong Qi}
\author{Shuang-Yin Huang}
\author{Chenghou Tu}
\author{Yongnan Li}
\email{liyongnan@nankai.edu.cn}
\affiliation{Key Laboratory of Weak-Light Nonlinear Photonics and School of Physics, Nankai University, Tianjin 300071, China}

\author{Wei Hu}
\author{Fei Xu}
\author{Yan-Qing Lu}
\affiliation{College of Engineering and Applied Sciences and National Laboratory of Solid State Microstructures, Nanjing University, Nanjing 210093, China}

\author{Hui-Tian Wang}
\email{htwang@nju.edu.cn}
\affiliation{Key Laboratory of Weak-Light Nonlinear Photonics and School of Physics, Nankai University, Tianjin 300071, China}
\affiliation{School of Physics and National Laboratory of Solid State Microstructures, Nanjing University, Nanjing 210093, China}
\affiliation{Collaborative Innovation Center of Advanced Microstructures, Nanjing University, Nanjing 210093, China}

\begin{abstract}
\noindent
Quantum protocols require access to large-scale entangled quantum states, due to the requirement of channel capacity. As a promising candidate, the high-dimensional orbital angular momentum (OAM) entangled states have been implemented, but only one of four OAM Bell states in each individual subspace can be distinguished. Here we demonstrate the first realization of complete OAM Bell-state measurement (OAM-BSM) in an individual subspace, by seeking the suitable unitary matrix performable using only linear optics and breaking the degeneracy of four OAM Bell states in ancillary polarization dimension. We further realize the superdense coding via our complete OAM-BSM with the average success probability of $\sim$82\% and the channel capacity of $\sim$1.1(4) bits. This work opens the window for increasing the channel capacity and extending the applications of OAM quantum states in quantum information in future.
\end{abstract}

%\pacs{ }

\maketitle
\end{CJK*}

%\begin{multicols}

Quantum protocols require access to large-scale entangled quantum states, due to the requirement of channel capacity \cite{R01}. As a promising candidate, the high-dimensional orbital angular momentum (OAM) entangled states have been implemented \cite{R02, R03, R04, R05, R06, R07, R08, R09, R10, R11}, but only one of four OAM Bell states in each individual subspace can be distinguished. A significant challenge is to resolve a complete Bell-state measurement, due to the requirement of channel capacity. The Bell-state measurement is required for quantum dense coding~\cite{R12, R13}, quantum teleportation~\cite{R14, R15}, quantum key distribution~\cite{R16, R17}, and entanglement swapping~\cite{R18,R19}. For polarization entangled Bell states, the complete Bell-state measurement using only linear optics and the classical communication with a 100\% efficiency is impossible \cite{R20, R21}. Although the complete polarization Bell-state measurement has realized by using nonlinear optics \cite{R22} or ancillary photons and linear optics \cite{R23}, they are inefficient and impractical in practical application. The polarization Bell-states with only linear optics have been distinguished completely in ancillary degree of freedom (DOF) \cite{R24,R25}, where the message is carried by the polarization and while the ancillary DOF is used to expand only the measurement space.

However, due to the limit of two dimensions, quantum protocols based on the polarization-entangled Bell states is limited to be 2 bits for a single qubit. Many quantum protocols require large-scale entangled quantum states \cite{R01}. The orbital angular momentum (OAM) carrying by the helical phase spatial mode \cite{R26} can generate a higher dimension Herbert space. Since the OAM entanglement was firstly realized \cite{R02}, it has attracted considerable attention in quantum optics \cite{R03, R04, R05, R06, R07, R08, R09}. The OAM quantum states should be a most promising candidate to provide access to the higher-dimension quantum states \cite{R10, R11} and then to increase the storage and processing potential of quantum information. However, only one of the four OAM Bell states can be distinguished from others in each individual subspace so far \cite{R01}, due to that the measurement results of three among four OAM Bell states are degenerate with the Hong-Ou-Mandel (HOM) interference. Therefore, the degeneracy of four OAM Bell states in each individual subspace must be broken to distinguish completely them and then to increase the channel capacity. 

Here we propose an idea for the first realization of complete OAM-BSM in each individual subspace with only linear optics and without auxiliary photons. The key of our idea is to break the degeneracy of four OAM Bell states in ancillary polarization dimension and seek the suitable unitary matrix performable using linear optics only. To confirm our idea, we finish the experimental verification for the complete OAM-BSM in an individual subspace with the quantum number of $m = 1$. We secondly realize the superdense coding by utilizing our complete OAM-BSM with the average success probability of $\sim$82\% and the channel capacity of $\sim$1.1(4).

Using the traditional method, we firstly generate the polarization-entangled photon states, and further produce the polarization-OAM entangled photon states similar to the method in Ref.~[24], so-called hyperentanglement. The introduction of polarization DOF is just used to expand the measurement space for breaking the degeneracy of OAM Bell states. When a fundamental Gaussian mode pumps a nonlinear $\beta$-BaBO$_3$ (BBO) crystal, such a kind of hyperentangled photon-pairs generated via the spontaneous parametric down-conversion (SPDC) can be described as

 \noindent
 \begin{align}\label{01}
|\Psi \rangle = \sum^{\infty}_{m=0}  c_m \lvert \Psi^{m+}\rangle \otimes \lvert \Psi^{s+} \rangle.
 \end{align}
Here $ c_m $ is a complex coefficient, $\lvert \Psi ^{s+} \rangle = (\lvert H \rangle_A \lvert V \rangle_B + \lvert V \rangle_A \lvert H \rangle_B) / \sqrt{2}$ is one of four polarization-entangled Bell states, $H$ $(V)$ represents the horizontal (vertical) polarization, and the subscripts $A$ and $B$ label the photon path. $\lvert \Psi^{m+} \rangle$ is one of four OAM Bell states in the $m$th-order subspace, which are written as 

\noindent
\begin{subequations}
\begin{align}\label{02}
|\Psi ^{m\pm} \rangle & = \dfrac{1}{\sqrt{2}} \left( \lvert +m \rangle_{A} \lvert-m \rangle_{B} \pm \lvert -m \rangle_{A} \lvert +m\rangle_{B} \right), \\
|\Phi^{m\pm} \rangle & = \dfrac{1}{\sqrt{2}} \left(\lvert +m \rangle_{A}\lvert +m \rangle_{B} \pm \lvert -m \rangle_{A} \lvert -m\rangle_{B} \right),
\end{align}
\end{subequations}
where $\lvert -m\rangle$ ($\lvert +m \rangle$) represents a photon state with an OAM of $-m \hbar$ ($+m \hbar$), while the case of $m=0$ is ignored.

\noindent
\begin{figure*}[!bhpt]
	\centerline{\includegraphics[width=13.5cm]{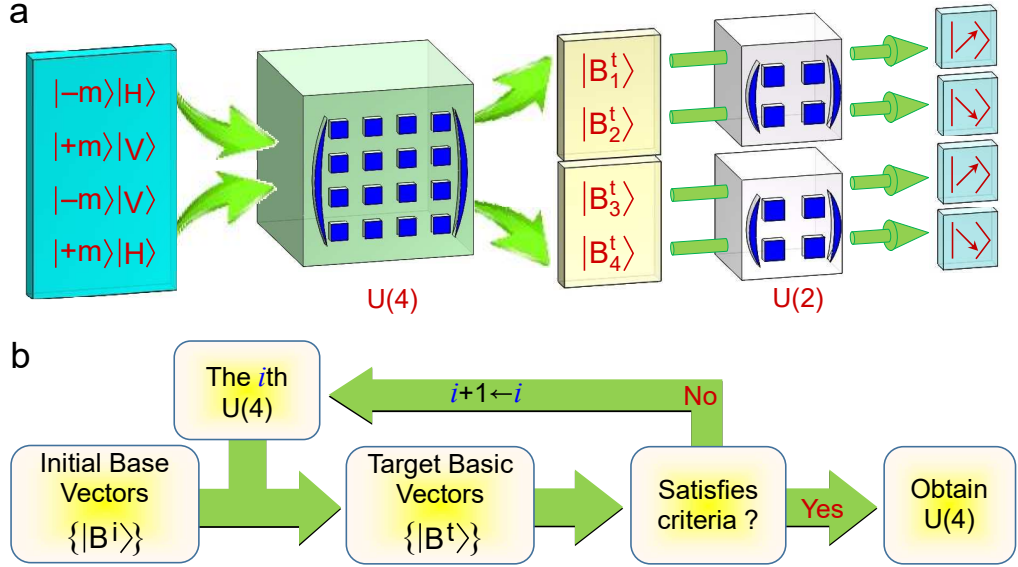}}
	\caption{Idea for the complete OAM-BSM based on the unitary transformation. \textbf{a}, Schematic diagram for the OAM-BSM. With the help of polarization DOF $\lvert \Psi^{s+} \rangle$, the complete OAM-BSM cannot be realized with the basis $\lvert -m \rangle \lvert H \rangle $, $\lvert +m \rangle \lvert V \rangle $, $\lvert - m \rangle \lvert V \rangle $, and $\lvert +m \rangle \lvert H \rangle$. After performing the unitary transformation $U(4)$, the target basis $\lvert B ^t_1 \rangle$, $\lvert B^t_2 \rangle$, $\lvert B^t_3 \rangle$, and $\lvert B^t_4 \rangle$ can be found, which can realize the complete OAM-BSM. Any one of two $U(2)$ unitary matrices transforms each target project basis vector into the fundamental Gaussian mode, in which photons can be detected easily. \textbf{b}, Principle of our model. First, finding out all of $U(4)$ under some reasonable condition. Then, use an $U(4)$ to calculate a set of target basis vectors. Finally, check whether the target basis vectors satisfies the criteria or not. If the criterion is not satisfied, the program starts again. While, if the criterion is satisfied, the $U(4)$ will be stored. Here the criterion is that in the suitable target basis vectors the coincidence measurements should be completely different for different OAM Bell states.} 
\end{figure*}

In an individual subspace, by applying one of four unitary operations on the OAM of one photon state $\lvert \Psi^{m+} \rangle \otimes \lvert \Psi^{s+} \rangle$, the other hyperentangled states can be obtained as follows. (1) with $\lvert \pm m \rangle_A$$\rightarrow$$\exp (\pm j \pi/2) \lvert \pm m \rangle_A$, which can be realized with a pair of Dove prisms oriented at an angle $ \pi/(4m) $ with respect to each other \cite{R01, R27}, $\lvert \Psi^{m+} \rangle \otimes \lvert \Psi^{s+} \rangle $$ \Rightarrow $$\lvert \Psi^{m-} \rangle \otimes \lvert \Psi^{s+} \rangle$; (2) with $\lvert +m \rangle_A$ $\leftrightarrow$$\lvert -m \rangle_A$, which can be realized by using a Dove prism, $ \lvert \Psi^{m+} \rangle \otimes \lvert \Psi^{s+} \rangle$$\Rightarrow$$\lvert \Phi^{m+} \rangle \otimes \lvert \Psi^{s+} \rangle $; (3) with $\lvert +m \rangle_A$$\leftrightarrow$$\lvert -m \rangle_A$ and $\lvert \pm m \rangle_A$$\rightarrow$$\exp (\pm j\pi/2) \lvert \pm m \rangle_A$ at the same time,  $\lvert \Psi^{m+} \rangle \otimes \lvert \Psi^{s+} \rangle$$\Rightarrow $$\lvert \Phi^{m-} \rangle \otimes \lvert \Psi^{s+} \rangle$ (see Fig.~4a for details). It should be noted that the Dove prism has no influence on the spin state of photons.
 
Here one of our two goals is to distinguish four OAM Bell states in an individual subspace with the aid of polarization entanglement. Combining Eqs.~(1) and (2), the four hyperentangled states $\lvert \Psi^{m+} \rangle \otimes \lvert \Psi^{s+} \rangle$, $\lvert \Psi^{m-} \rangle \otimes \lvert \Psi^{s+} \rangle$, $\lvert \Phi^{m+} \rangle \otimes \lvert \Psi^{s+} \rangle$, and $\lvert \Phi^{m-} \rangle \otimes \lvert \Psi^{s+} \rangle$ can be written as a superposition of the basis $\{ \lvert B^i_k \rangle_A \lvert B^i_l \rangle_B \}$ (see Supplementary Information for details), where $k,l\in \{1,2,3,4\} $ and $\lvert B^i_1 \rangle = \lvert -m \rangle \lvert H \rangle $, $\lvert B^i_2 \rangle = \lvert +m \rangle \lvert V \rangle$, $\lvert B^i_3 \rangle = \lvert -m \rangle \lvert V \rangle $, and $\lvert B^i_4 \rangle = \lvert +m \rangle \lvert H \rangle $. After projecting the four photon-pair hyperentangled states into the basis $\{\lvert B^i_k \rangle\}$ with $k = \{1,2,3,4\}$ and performing the coincidence measurement between photons A and B, the results are the same for $\lvert \Psi^{m+} \rangle \otimes \lvert \Psi^{s+} \rangle $ and $\lvert \Psi^{m-} \rangle \otimes \lvert \Psi^{s+} \rangle $ ($ \lvert \Phi^{m+} \rangle \otimes \lvert \Psi^{s+} \rangle$ and $\lvert \Phi^{m-} \rangle \otimes \lvert \Psi^{s+} \rangle$) (Supplementary Fig.~S1). That is to say, although there has the assistance of polarization DOF $\lvert \Psi^{s+} \rangle$, four OAM Bell states $\lvert \Psi^{m+} \rangle$, $\lvert \Psi^{m-} \rangle$, $\lvert \Phi^{m+} \rangle$ and $\lvert \Phi^{m-} \rangle$ cannot still be distinguished completely under the basis $\{\lvert B^i \rangle\}$ measurement. Therefore, a key step is to transform the initial basis $\{\lvert B^i_k \rangle\}$ with $k = \{1,2,3,4\}$ into a suitable project basis $\{\lvert B^t_k \rangle\}$ with $k = \{1,2,3,4\}$, which need to find out a suitable unitary transformation matrix $U(4)$ (Fig.~1a for idea). This suitable project basis is called the target basis here.

Inspired by Krenn \textit{et al.}~\cite{R28}, we construct a theoretical model to seek a target  basis vectors (Fig.~1). Our calculation results demonstrate that there are a series of unitary matrices $U(4)$, in other words, the unitary matrices $U(4)$ satisfying our criterion are not the only one. We use one of them, as shown in Eq.~(3) below, to guide our experiment 
\begin{align}\label{eq3}
U(4) = \dfrac{1}{\sqrt{2}}
\left( \begin{array}{cccc}
1 &  1 & 0 & 0 \\
1 & -1 & 0 & 0 \\
0 &  0 & 1 & 1 \\
0 &  0 & 1 & -1 \\ \end{array} \right).
\end{align}
Then the corresponding target basis is
\begin{align}\label{eq4}
\left( \begin{array}{cccc}
\lvert B^t_1 \rangle \\
\lvert B^t_2 \rangle \\
\lvert B^t_3 \rangle \\
\lvert B^t_4 \rangle \end{array} \right)
= \dfrac{1}{\sqrt{2}} 
\left( \begin{array}{cccc}
\lvert -m \rangle \lvert H \rangle + \lvert +m \rangle \lvert V \rangle  \\
\lvert -m \rangle \lvert H \rangle -  \lvert +m \rangle \lvert V \rangle  \\
\lvert -m \rangle \lvert V \rangle + \lvert +m \rangle \lvert H \rangle  \\
\lvert -m \rangle \lvert V \rangle -  \lvert +m \rangle \lvert H \rangle  \end{array} \right).
\end{align}
With this target basis vector, the four hyperentangled states become into

\noindent
\begin{subequations}
\begin{align}\label{05}
 \lvert \Psi^{m+} \rangle \otimes \lvert \Psi^{s+} \rangle = & + \dfrac{1}{2} \left( \lvert B^t_1 \rangle_A \lvert B^t_1 \rangle_B - \lvert B^t_2 \rangle_A \lvert B^t_2 \rangle_B \right.  \nonumber \\
 & + \left. \lvert B^t_3 \rangle_A \lvert B^t_3 \rangle_B - \lvert B^t_4 \rangle_A \lvert B^t_4 \rangle_B \right), \\
  \lvert \Psi^{m-} \rangle \otimes \lvert \Psi^{s+} \rangle = & + \dfrac{1}{2} \left( \lvert B^t_2 \rangle_A \lvert B^t_1 \rangle_B - \lvert B^t_1 \rangle_A \lvert B^t_2 \rangle_B \right. \nonumber \\
 & + \left. \lvert B^t_4 \rangle_A \lvert B^t_3 \rangle_B - \lvert B^t_3 \rangle_A \lvert B^t_4 \rangle_B \right), \\
 \lvert \Phi^{m+} \rangle \otimes \lvert \Psi^{s+}\rangle = & + \dfrac{1}{2} \left( \lvert B^t_3 \rangle_A \lvert B^t_1 \rangle_B + \lvert B^t_4 \rangle_A \lvert B^t_2 \rangle_B \right.  \nonumber \\
 & + \left. \lvert B^t_1 \rangle_A \lvert B^t_3 \rangle_B + \lvert B^t_2 \rangle_A \lvert B^t_4 \rangle_B \right), \\ 
 \lvert \Phi^{m-}\rangle \otimes \lvert \Psi^{s+} \rangle = & - \dfrac{1}{2} \left( \lvert B^t_4 \rangle_A \lvert B^t_1 \rangle_B + \lvert B^t_3 \rangle_A \lvert B^t_2 \rangle_B \right. \nonumber \\
 & + \left. \lvert B^t_2 \rangle_A \lvert B^t_3 \rangle_B + \lvert B^t_1 \rangle_A \lvert B^t_4 \rangle_B \right).
\end{align}
\end{subequations}

\begin{figure*}[bht]
	\centerline{\includegraphics[width=17.5cm]{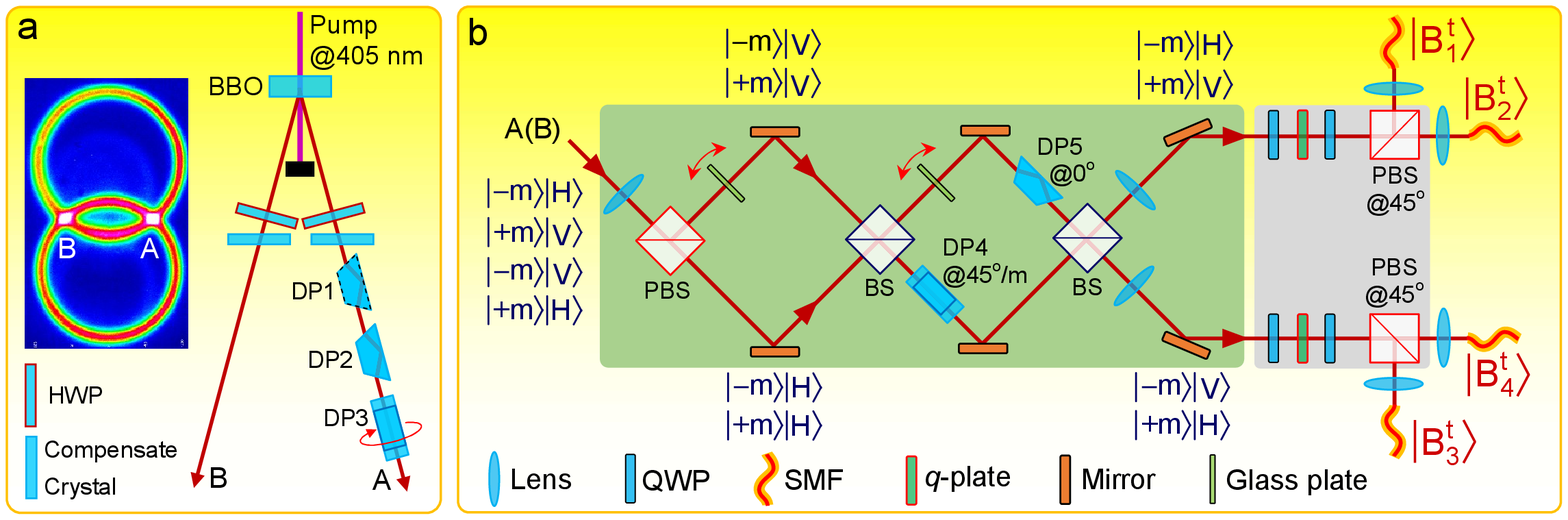}}
	\caption{Preparation and analysis of the OAM Bell states. \textbf{a}, The preparation of the source of the hyperentanglement state. A half wave plate (HWP) and a compensate crystal in each path are used to compensate the work-off effects. \textbf{b}, the experimental setup of the OAM Bell-state analyzer. Photon A (B) of an OAM Bell state is separated by a PBS according to its polarization. Then a modified Mach-Zehnder interferometer (MZI) makes the photon in the state $\lvert -m \rangle \lvert H \rangle $ or $\lvert +m \rangle \lvert V \rangle $ ($\lvert -m \rangle \lvert V \rangle $ or $\lvert +m \rangle \lvert H \rangle $) exit on the top (bottom) output port of the second BS. Actually, the PBS and the MZI play the part of the unitary matrix $U(4)$ in Fig.~1(a). The role of the \textit{q}-plate (sandwiched by two quarter wave plates) and the PBS@45$^\circ$ together is the unitary matrix $U (2)$ in Fig.~1(a). An interference filter (not be shown) with a 3-nm bandwidth centred at 810 nm is used to remove the unwanted photons in front of each single mode fiber (SMF).}
\end{figure*}

In fact, there are sixteen combination basis states of photons A and B ($\{ \lvert B^t_k \rangle_A \lvert B^t_l \rangle_B \}$ with $k, l \in \{1,2,3,4\}$). We can see from Eq.~(5) that each hyperentangled state is a unique superposition of four among the sixteen possible combination target basis states. By carrying out the local Bell-state measurement, the four OAM Bell states in an individual subspace can be distinguished completely. 

In experiment, the hyperentangled photon-pairs source is generated by pumping a 2-mm-thick BBO nonlinear crystal (see Fig.~2a). The pump light is a femtosecond pulsed laser with a power of 230 mW, a pulse duration of $\sim$140 fs, a repetition rate of $\sim$80 MHz, and a central wavelength of 405 nm. Under the type-II phase matching, the down-converted photons at a degenerate wavelength of 810 nm are emitted into two cones. Only the photons located at the two overlapping areas of two cones (inset of Fig.~2a) are useful. Three Dove prisms (DP1, DP2, and DP3) are used to transform the OAM Bell states with each other. The pump light is a fundamental Gaussian beam with a beam waist of $\omega_0 \sim 0.9$ mm. The beam waist locates at the center plane of the BBO crystal.

The photons in each path will meet a polarization beam splitter (PBS), which makes the photons of different polarization enter into the first beam splitter (BS) from different input ports, and a glass plate is used to control the equal optical paths of two arms between the PBS and the first BS (Fig.~2b). Two BSs form a modified Mach-Zehnder interferometer (MZI), which contains a pair of Dove prisms oriented at an angle of $\pi/(4m)$ with respect to each other (Fig.~2b). The difference of optical paths between two arms of the MZI is adjusted by another glass plates to make the states $ \lvert -m \rangle \lvert H \rangle$ and $ \lvert -m \rangle \lvert V \rangle$ ($ \lvert +m \rangle \lvert V \rangle$ and $\lvert +m \rangle \lvert H \rangle$) emit from different output ports of the second BS (Fig.~2b). In fact, the PBS and the modified MZI together finish the function of the unitary matrix $U(4)$. Then performing a $U(2)$ transform, which is composed of a $m/2$-order \textit{q}-plate~\cite{R29,R30} sandwiched between two quarter wave plates (QWPs), the states $\lvert B^t_1\rangle$ (or $\lvert B^t_3 \rangle$) and $\lvert B^t_2 \rangle$ (or $\lvert B^t_4 \rangle$) can be converted into $\lvert \nearrow \rangle$ and $\lvert \searrow \rangle$, respectively (Fig.~1a and Fig.~2b). Here $\lvert \nearrow \rangle \propto (\lvert H \rangle + \lvert V \rangle)$ and $\lvert \searrow \rangle \propto (\lvert H \rangle - \lvert V \rangle) $ can be separated with a PBS$@$45$^\circ$. It should be noted that between the BBO crystal and the \textit{q}-plate, a 4f system composed of two lenses images the center plane of the BBO crystal onto the \textit{q}-plate. Photons in the four target basis states $|B^t \rangle$ exit from different output ports of two PBSs, respectively (Fig.~2b). A single mode fiber (SMF) is utilized to collect the photons in the state $\lvert \nearrow \rangle $ or $\lvert \searrow \rangle $ and filter others. The sixteen possible coincidence measurement results between the paths A and B can be carried out under the target project basis vectors.

The coincidence measurement for $m = 1$ has been carried out to verify our idea and experimental scheme (Fig.~3). For any one of the four OAM Bell states, only a group of unique four combinations give the coincidence measurement signals. Therefore, the four OAM Bell states can be distinguished completely. If one of the combination states $\lvert B^t_k \rangle_A \lvert B^t_k \rangle_B$ with $k = \{1,2,3,4\}$ is fired, the input OAM Bell state must be $ \lvert \Psi ^{1+} \rangle$. Similarly, for $ \lvert \Psi ^{1-} \rangle$, the combination $\lvert B^t_2 \rangle_A \lvert B^t_1 \rangle_B$, $\lvert B^t_1 \rangle_A \lvert B^t_2 \rangle_B$, $\lvert B^t_4 \rangle_A \lvert B^t_3 \rangle_B$ or $\lvert B^t_3 \rangle_A \lvert B^t_4 \rangle_B$ should give the coincidence signals. And $\lvert B^t_3 \rangle_A \lvert B^t_1 \rangle_B$, $\lvert B^t_4 \rangle_A \lvert B^t_2 \rangle_B$, $\lvert B^t_1 \rangle_A \lvert B^t_3 \rangle_B$, or $\lvert B^t_2 \rangle_A \lvert B^t_4 \rangle_B$ ($\lvert B^t_4 \rangle_A \lvert B^t_1 \rangle_B$, $\lvert B^t_3 \rangle_A \lvert B^t_2 \rangle_B$, $\lvert B^t_2 \rangle_A \lvert B^t_3 \rangle_B$, $\lvert B^t_1 \rangle_A \lvert B^t_4 \rangle_B$ ) has coincidence for the state $ \lvert \Phi ^{1+} \rangle$ ($ \lvert \Phi ^{1-} \rangle$). In our experiment, the signal-noise ratio (SNR) of the state, which is defined as the ratio between the sum of four ratios of the actual states and the sum of the other twelve ratios, are ${\rm SNR}_{\lvert \Psi ^{1+} \rangle} = 6.78$, ${\rm SNR}_{\lvert \Psi ^{1-} \rangle} = 4.6$, ${\rm SNR}_{\lvert \Phi ^{1+} \rangle} = 5.09$, and ${\rm SNR}_{\lvert \Phi ^{1-} \rangle} = 3.12$. 

\begin{figure*}[bht]
	\centerline{\includegraphics[width=17.5cm]{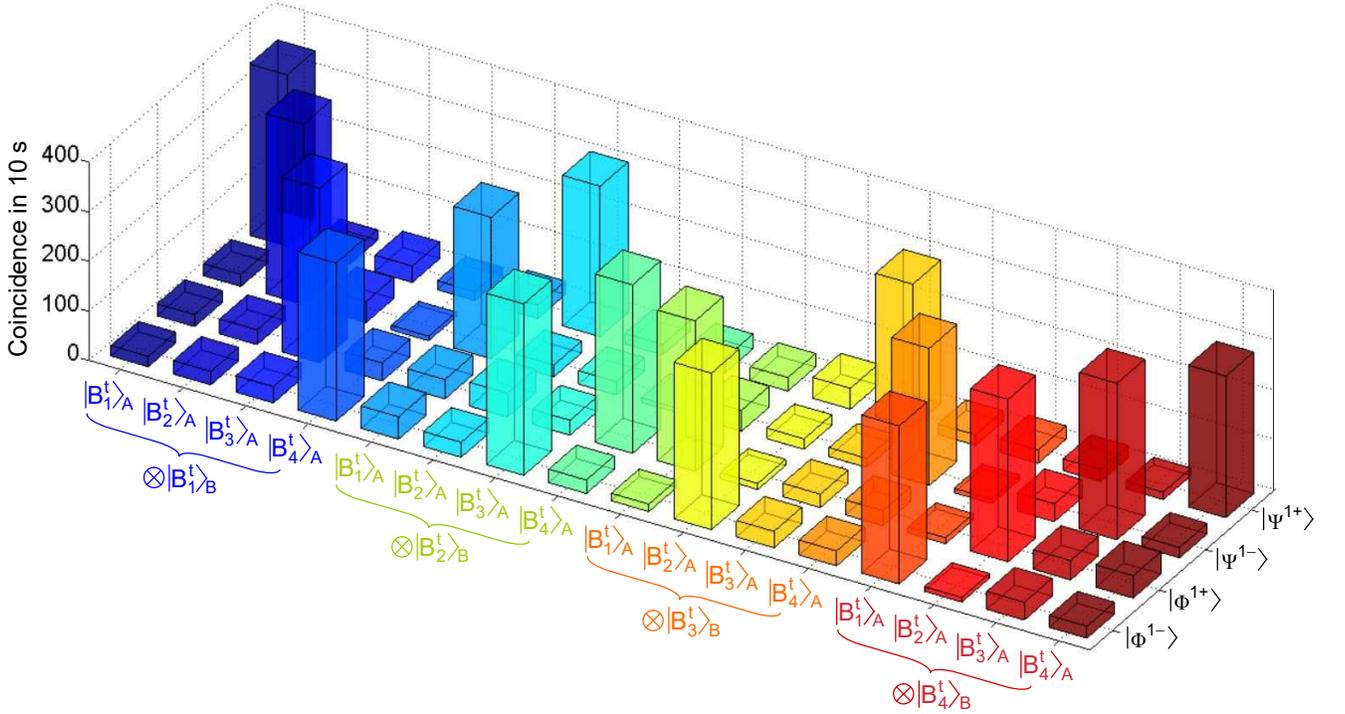}}
	\caption{Experimental results of OAM-BSM for $m = 1$. The vertical axis represents the coincidence counts in 10 seconds.} 
\end{figure*}

\begin{figure*}[bht]
	\centerline{\includegraphics[width=17.5cm]{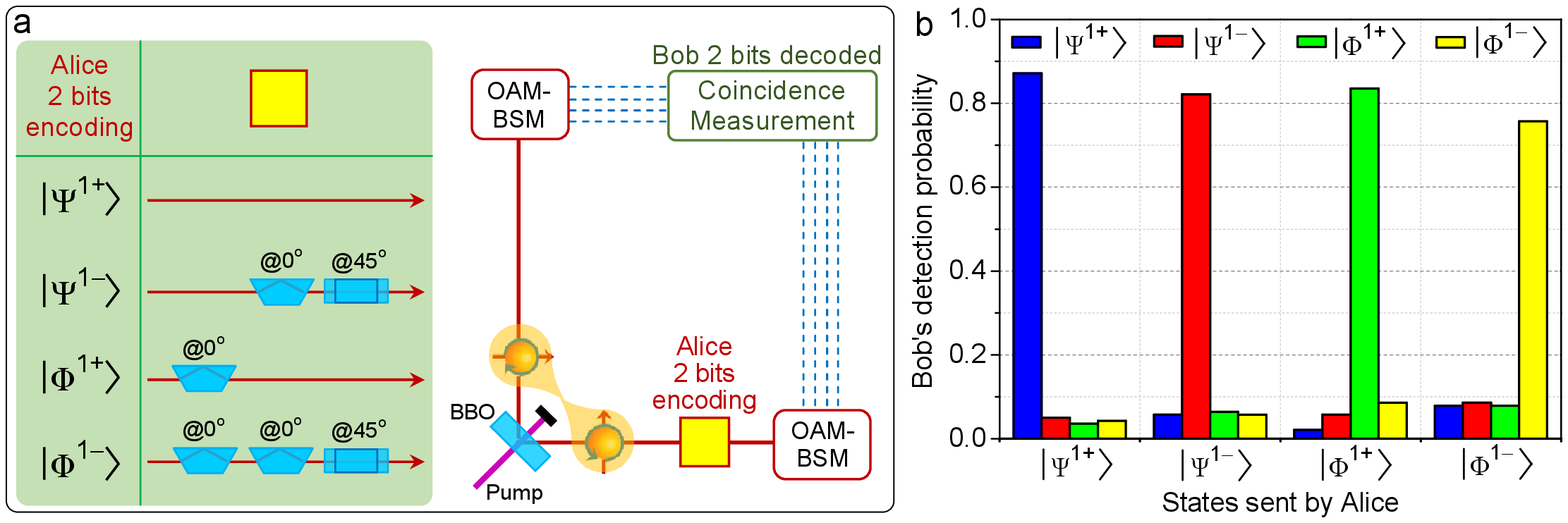}}
	\caption{ \textbf{a}, Experimental setup for superdense coding. Alice encodes two bits on her photon by using three Dove prisms to transform the OAM Bell-state shown in the table. Bob performs the OAM-BSM and decodes two bits with coincidence measurement. \textbf{b}, Bob's experimental detection probability for superdense coding. Bob infers the state encoded and sent by Alice with the probabilities as vertical axis shown. His average success probability is about 82\%.}
\end{figure*}

Based on our OAM-BSM, a superdense-coding protocol can be realized (Fig.~4a). The hyperentanglement source is produced via the SPDC in a nonlinear BBO crystal (as described above). Photon pairs are entangled in OAM, polarization, and emission time synchronously \cite{R31}. Alice encodes two-bits message carried by the OAM Bell state, by operating her photon with three Dove prisms properly (Fig.~4a). Then Bob performs the OAM-BSM and decodes the two-bits message sent by Alice with the OAM-BSM. Our superdense-coding implementation has been characterized, by switching between the four states and measuring the output states. As the coincidence counts for each input state (Fig.~4b), due to the imperfections in the optical elements, alignment, and input states, the average success probability is $\sim$82\%. The channel capacity is calculated by

\begin{align}\label{eq6}
{\rm CC} = \mathop \mathrm{max} \limits_{p(\textbf{x})} \sum^{4}_{x,y=1} p(y|x) p(x)  \log \dfrac{p(y|x)}{\mathop \sum\limits^{4}_{x'=1} p(y|x') p(x') }.
\end{align}
\noindent
Here $ p(y|x) $ represents the conditional probabilities when the state \textbf{x} is sent by Alice and the state \textbf{y} is decoded by Bob. $p(\textbf{x}) =\{ p(\Psi^{1+}), p(\Psi^{1-}), p(\Phi^{1+}), p(\Phi^{1-})\}$ maximizes the capacity. In our experiment, for a uniform probability of transmission, ${\rm CC} = 1.1(4)$.

We firstly resolved an important problem---the complete OAM-BSM, which is the requirement of channel capacity in quantum information. Since the limit of the two dimensions, the channel capacity of the polarization entanglement can never be higher than two bits. To break the limit of two-bits channel capacity, the OAM entanglement is a good candidate, because the OAM states can construct an infinite dimension Herbert space. So far, however,only one of the four OAM Bell states in an individual subspace can be distinguished from others with the HOM interference. The realization of a complete OAM-BSM is still a crucial challenge. The key of our solution is to break completely the degeneracy of four OAM Bell states, by introducing the polarization-OAM hyperentanglement and performing the suitable unitary transform with only linear optics. We propose a theoretical model to seek the suitable unitary matrix $U(4)$, which can transform the initial basis \{$\lvert B^i \rangle$\} into the target basis \{$\lvert B^t \rangle$\}. In particular, the unitary transformation can be performed experimentally using only linear optics. We verified experimentally the efficient and practical realization of the complete OAM-BSM in the individual OAM subspace of $m=1$. We secondly realized the superdense coding by utilizing our complete OAM-BSM, and the results show that the average success probability is $\sim$82\% and the channel capacity is $\sim$1.1(4). Since the OAM states can generate an infinitely dimensional discrete Hilbert space, the channel capacity can be higher than 2 by encoding the message. Although our experimental results does not reach this goal at present, our research opens the window for increase of the channel capacity and extend the applications of photon OAM states in quantum information science in future.

\noindent {\bf Acknowledgements}

\noindent This work was supported by the National key R\&D Program of China under grant Nos. 2017YFA0303800 and 2017YFA0303700, and the National Natural Science Foundation of China under grant Nos. 11534006 and 11374166. We acknowledge the support by Collaborative Innovation Center of Extreme Optics. The authors thank C. F. Li for helpful discussions and valuable advice.

\end{document}